\documentclass[reprint,amsmath,amssymb,aps,prb]{revtex4-1}
\usepackage{graphicx}
\usepackage[colorlinks=true, citecolor=blue, linkcolor=red]{hyperref}
\usepackage{bm}
\setcitestyle{numbers,square}
\usepackage{accents}

\newcommand{\ut}[1]{\underaccent{\tilde}{#1}}

\begin{document}

\preprint{APS/123-QED}

\title{Double Dirac Cones and Topologically Non-Trivial Phonons for Continuous, Square Symmetric (\texorpdfstring{C$_{4v}$}{Lg} and \texorpdfstring{C$_{2v}$}{Lg}) Unit Cells}

\author{Yan Lu}
\affiliation{Department of Mechanical Engineering, Boston University, Boston, MA 02215}

\author{Harold S. Park}
\email[Corresponding author: ]{parkhs@bu.edu}
\affiliation{Department of Mechanical Engineering, Boston University, Boston, MA 
02215}

\date{\today}

\begin{abstract}
Because phononic topological insulators have primarily been studied in discrete, graphene-like structures with C$_{6}$ or C$_{3}$ hexagonal symmetry, an open question is how to systematically achieve double Dirac cones and topologically non-trivial structures using continuous, non-hexagonal unit cells.  Here, we address this challenge by presenting a novel computational methodology for the inverse design of continuous two-dimensional square phononic metamaterials exhibiting C$_{4v}$ and C$_{2v}$ symmetry.  This leads to the systematic design of square unit cell topologies exhibiting a double Dirac degeneracy, which enables topologically-protected interface propagation based on the quantum spin Hall effect (QSHE).  Numerical simulations prove that helical edge states emerge at the interface between two topologically distinct square phononic metamaterials, which opens the possibility of QSHE-based pseudospin-dependent transport beyond hexagonal lattices.
\end{abstract}

\maketitle

The discovery of topological insulators (TIs) in quantum mechanical systems \cite{Haldane,Zhang,Kane,hazanRMP2010,huberNP2016,wangNM2017} has stimulated significant interest in developing topological analogues in other fields, including acoustics~\cite{cummerNRM2016,yangPRL2015}, photonics~\cite{khanikaevNP2017,luNP2014,ozawaRMP2019}, and phononics~\cite{chaNATURE2018,dongNM2017,huberNP2016}.  TIs have attracted significant scientific interest primarily due to their potential to enable lossless propagation of wave energy along well-defined interfaces~\cite{mooreNATURE2010,heNP2016,qiPT2010}.  There are various approaches to obtaining topologically-protected wave propagation, most of which are based on either the quantum hall effect (QHE)~\cite{haldanePRL1988}, the quantum spin hall effect (QSHE)~\cite{qiPT2010}, or the quantum valley hall effect (QVHE)~\cite{renRPP2016}.  Many reports on obtaining topologically-protected wave propagation have emerged in recent years using these fundamental principles~\cite{mousaviNC2015,huberNP2016,heNP2016,nashPNAS2015,wangPRL2015,yangPRL2015,khanikaevNC2015,susstrunkPNAS2016,susstrunkSCIENCE2015,palJAP2016,chaNATURE2018,dongNM2017,jiangNANOSCALE2018,liNC2018,liuPRA2018,miniaciPRX2018,wuSR2018,yanNM2018,yuNC2018,zhuPRB2018,chenMSSP2021,duJMPS2020,saba2020nature}. 

A commonality to nearly all studies on phononic TIs is their dependence on using graphene-like discrete, hexagonally-symmetric (i.e. C$_{6}$ or C$_{3}$) lattice structures to obtain double Dirac cones as the first step to achieving topologically non-trivial structures.  While the usage of hexagonal symmetries is well-established, this limits the design space for TIs, and fundamental questions regarding the structure and resulting topological properties of different structural symmetries remain unresolved.  Because of this, researchers have recently investigated the possibility of achieving Dirac cones in non-hexagonal lattices~\cite{miertPRB2016,qinPCCP2020,luoNL2015,ezawaNJP2014,xiaPRB2018,li2019dual,huo2019edge}.  However, the challenge remains to create continuous unit cells that do not rely on hexagonal symmetry in order to exhibit a double Dirac cone, while also forming the basis for spin orbit coupling-based mode inversion, thus enabling QSHE-based topologically protected interfacial wave propagation.

In this letter, we address this challenge by presenting a novel multi-objective gradient-based topology optimization approach that enables, in contrast to previous topology optimization methods for TIs~\cite{nanthaJMPS2019,christiansenPRL2019,chenMSSP2021}, the ability to design continuous, square symmetric unit cells that serve as the building blocks for phononic TIs. We utilize the design procedure to obtain a double Dirac cone using anti-plane shear waves in an elastic square lattice, and show that breaking symmetry of the resulting square topology leads to a topologically non-trivial bandgap and pseudospin-dependent wave propagation. Therefore, this work opens the possibility of QSHE-based pseudospin-dependent transport beyond hexagonal lattices.

\begin{figure*}
\includegraphics[width=\textwidth]{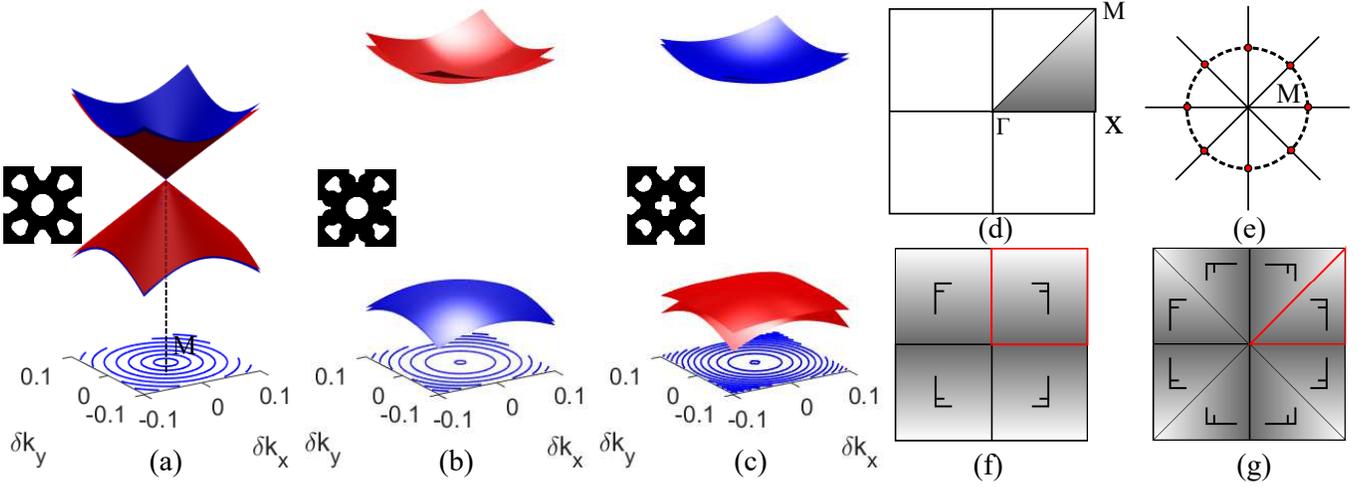}
\caption{Optimized C$_{4v}$ unit cells with (a) Double Dirac cones, (b) complete band gap, (c) mode inversion with respect to (b). The square lattice consists 120$\times$120 pixels, where each pixel stands for aluminum (black) or void (white).  The eigenvalue surfaces are calculated using the finite element method around the high symmetry point $M$. The surfaces are colored in red and blue which stand for different mode symmetries. The inverted surface colors in (b) and (c) implies spin orbit coupling-induced mode inversion. The equi-frequency contours correspond to the eigenvalue surfaces colored in blue. (d) First Brillouin zone of a square unit cell. The irreducible Brillouin zone is shaded in gray. (e) Frequency constraint points around $M$. Illustration of (f) C$_{2v}$ and (g) C$_{4v}$ symmetry. The minimal design region is enclosed by red line segments.}\label{Fig:schematic}
\end{figure*}

Our approach is a three-stage design strategy to match the eigensystem properties of the Bernevig-Hughes-Zhang (BHZ) effective Hamiltonian~\cite{bernevigSCIENCE2006} described in detail in the SI~\cite{SI}, in which the frequency of the double Dirac degeneracy and the width of the topological band gap are user-specified values.   We first search for square lattices with a four-fold (double) Dirac degeneracy, as illustrated by the C$_{4v}$ unit cell in Fig. \ref{Fig:schematic}(a).  Next, we lift the quadruple degeneracy to open a complete band gap as in Fig. \ref{Fig:schematic}(b). Finally, we induce mode inversion as in Fig. \ref{Fig:schematic}(c) such that the interface of the two unit cells in Figs. \ref{Fig:schematic}(b,c) supports topologically protected wave propagation. 

Each design stage leading to the continuous elastic square unit cells brings unique challenges, which we address through different objective functions governing the topology optimization.  In the first stage when obtaining the double Dirac degeneracy, Herring's rule \cite{herring1937effect} on irreducible representations specifies that in hexagonal unit cells with C$_{6v}$ symmetry, the double Dirac degeneracy consists of two deterministic degeneracies, which can easily be brought together to create the double degeneracy~\cite{mousaviNC2015}. In contrast, the double Dirac degeneracy for C$_{4v}$ and C$_{4}$ symmetries consists of mixed accidental and deterministic degeneracies, while the double Dirac degeneracy for C$_{2v}$ and C$_{2}$ symmetries can only consist of accidental degeneracies. Because it is non-trivial to induce double Dirac degeneracies between deterministic and accidental, or two accidental degeneracies, we leverage topology optimization to accomplish this via the first set of objective functions to a design frequency $\omega_0$ 
\begin{equation}
h=|\omega_0-\omega(\mathbf{k_0})_i|,
\label{Eq:Objquadrupledegen}
\end{equation}
where $\omega(\mathbf{k_0})_i$ is the frequency of $i$th band at the high symmetry point $\mathbf{k_0}$, or the $M$ point in our approach. 

A further challenge when creating spin-degenerate structures as in Figs. \ref{Fig:schematic}(a-c) is to enforce the coalescence of eigenvalue surfaces that is inherent in the BHZ model.  While relatively simple for hexagonal lattices, it is non-trivial for square lattices at the high symmetry points of the irreducible Brillouin zone~\cite{li2019dual,ezawaNJP2014,huo2019edge}, where failure to match the eigenvalue surfaces along with the resulting small and anisotropic group velocities leads to significant loss along the topological interface~\cite{christiansenPRL2019,dong2020customizing,chenMSSP2021}. To overcome this challenge, we assign eight frequency constraint points along the high symmetry axis of the first Brillouin zone as in Fig. \ref{Fig:schematic}(e), forming a circle which encloses the high symmetry ($M$) point of interest. This design objective constrains the equi-frequency contours to be circular as in Figs. \ref{Fig:schematic}(a-c), while satisfying that the eigenvalues of the BHZ Hamiltonian depend upon the distance away from the degeneracy, $|\delta k|=\sqrt{\delta k_x^2+\delta k_y^2}$. The frequency constraints bring the maximum and minimum frequencies along the circle for each of the two phonon bands that will be degenerate above and below the design frequency $\omega_0$ to the same frequency value.  The reason for this is two-fold:  it results in matching not only of the eigenvalue surfaces, as shown in Fig. \ref{Fig:schematic}(a) and the surfaces colored in blue in Figs. \ref{Fig:schematic}(b,c), but also isotropy and matching of the group velocity near the high symmetry ($M$) point, which is required in QSHE-based TIs~\cite{mousaviNC2015}.

\begin{figure*}
\center
\includegraphics[width=\textwidth]{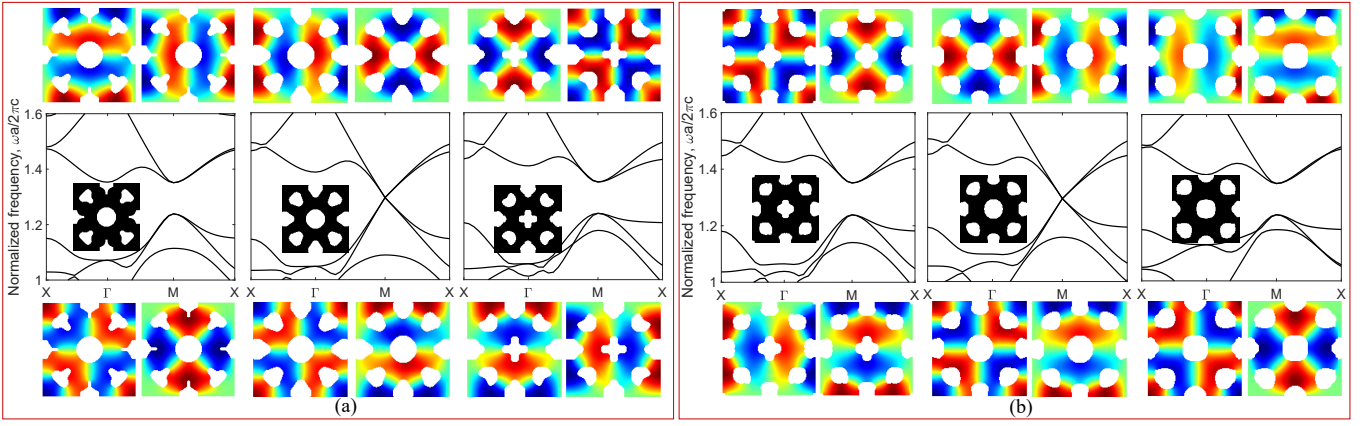}
\caption{(a) Optimized C$_{4v}$ unit cells that exhibit double Dirac cones (middle), band gap opening in second stage of optimization (left) and mode inversion (right).(b) Optimized C$_{2v}$ unit cells that exhibiting double Dirac cones (middle), band gap opening in second stage of optimization (left) and mode inversion (right).}\label{Fig:C4v_C2v}
\end{figure*}  
For the eigenvalue surfaces that are colored in red in Figs. \ref{Fig:schematic}(b,c), some band splitting along $\Gamma$-$M$ is inevitable due to the compatibility relations between different symmetries~\cite{dresselhaus2007group}. The circular frequency constraints in this case serve to minimize the splitting between the branches. Numerical experiments have shown that these frequency constraints can be satisfied by the 6th to 9th bands at the $M$ point. These frequency constraints can also be written as objective functions which minimize the aforementioned frequency differences (see the SI for details~\cite{SI}).

Another issue to resolve emerges when opening a topologically non-trivial band gap from the initial quadruple degeneracy.  Specifically, in a square lattice where spin states may potentially be formed entirely or in-part by accidental degeneracies, a risk that arises in opening a band gap from the double Dirac degeneracy is the corresponding lifting of the accidental degeneracy.  To ensure that the accidental degeneracy remains after the band gap is opened, we assign two degenerate frequencies, one below ($\ut{\omega}_0$) and the other above ($\tilde{\omega}_0$) the user-defined topological band gap. Therefore, the corresponding objective functions for the corresponding $i$th mode become 
\begin{align}
\label{Eq:Objbandopeninglower}
h=|\ut{\omega}_{0}-\omega(\mathbf{k_0})_i|;\; i=\ut{i}_1,\ut{i}_2,\\
\label{Eq:Objbandopeningupper}
h=|\tilde{\omega}_{0}-\omega(\mathbf{k_0})_i|;\; i=\tilde{i}_1,\tilde{i}_2,
\end{align}
which, in conjunction with circular frequency constraints, results in structures and band gaps as in Fig. \ref{Fig:schematic}(b).  

Once the topology that has a band gap with two sets of degenerate modes at the top (modes $\tilde{i}$) and bottom of (modes $\ut{i}$) the band gap is obtained, the resulting modeshapes of the degenerate modes are used as the reference modes for the subsequent mode inversion step, which mimics strong spin orbit coupling (SOC)~\cite{mousaviNC2015}. The quality of mode inversion is measured by the square of normalized inner product of the displacement field, also known as the modal assurance criterion (MAC)\cite{kim2000mac}
\begin{equation}\label{Eq:MAC}
\gamma_{i,j}=\dfrac{\langle U_i^{0}|U_j\rangle^2}{\parallel U_i^{0}\parallel^2 \parallel U_j\parallel^2},
\end{equation}
where $U_i^0$ denotes the $i$th reference displacement mode, and $U_j$ denotes the $j$th displacement mode in the unit cell with mode inversion. The value of $\gamma_{i,j}$ varies between 0 and 1 such that when $\gamma_{i,j}$ is equal to one, the modeshapes of $U_i^0$ and $U_j$ are the same. 

The optimization process to achieve mode inversion includes two steps. It starts from the four degenerate modes $j$ in the double Dirac degeneracy as in Fig. \ref{Fig:schematic}(a), where we identify $\ut{j}$ and $\tilde{j}$ by searching for $\max\lbrace\gamma_{i,j}\rbrace$ given $\ut{i}$ and $\tilde{i}$, respectively. $\max\lbrace\gamma_{i,j}\rbrace$ between the corresponding modes is then maximized to improve the inversion quality. Simultaneously, the frequency differences between $\ut{\omega}_0$ and $\omega(\mathbf{k_0})_{\tilde{j}}$, $\tilde{\omega}_0$ and $\omega(\mathbf{k_0})_{\ut{j}}$ are minimized to induce mode inversion by bringing modes $\tilde{j}$ to the lower edge of the band gap, and modes $\ut{j}$ to the upper edge of the band gap.  This mode inversion process is achieved through the following objective functions
\begin{align}
\label{Eq:ObjMAC}
\nonumber f&=-\max\lbrace\gamma_{i,j}\rbrace;\; i=\ut{i}_1,\ut{i}_2,\tilde{i}_1,\tilde{i}_2\\
j&=\ut{j}_1,\ut{j}_2,\tilde{j}_1,\tilde{j}_2\\
\label{Eq:Objmodeinversionlower}
h&=|\ut{\omega}_{0}-\omega(\mathbf{k_0})_{\tilde{j}}|,\; \tilde{j}=\tilde{j}_1,\tilde{j}_2\\
\label{Eq:Objmodeinversionupper}
h&=|\tilde{\omega}_{0}-\omega(\mathbf{k_0})_{\ut{j}}|,\;\ut{j}=\ut{j}_1,\ut{j}_2
\end{align} 
which leads to structures and band gaps as in Fig. \ref{Fig:schematic}(c). We emphasize that we do not need to explicitly prescribe symmetry breaking to mimic SOC in order to successfully complete this stage of the optimization process.

\begin{figure*}
\center
\includegraphics[width=\textwidth]{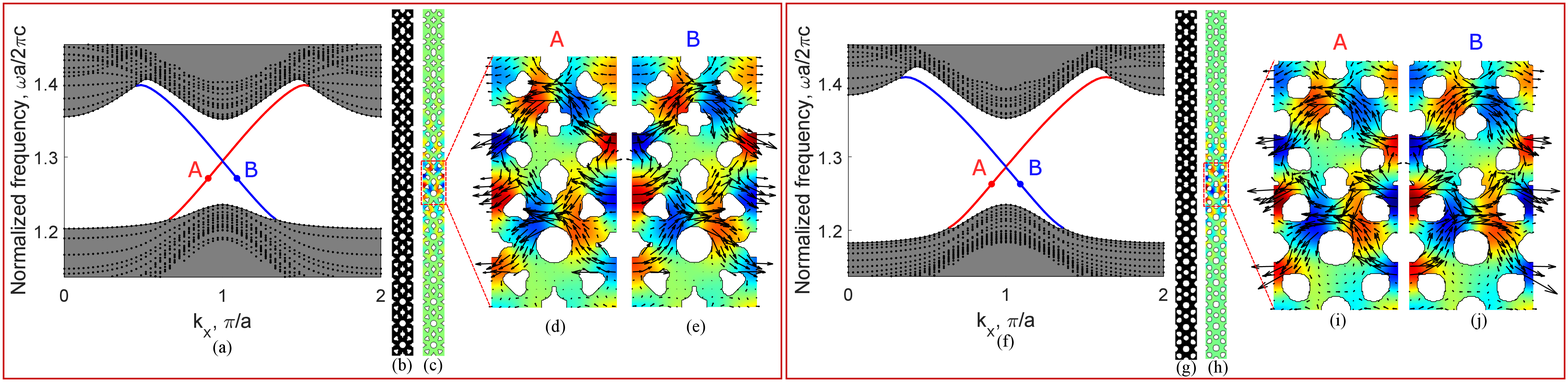}
\caption{Band structure calculations for supercell constructed from (left) C$_{4v}$ and (right) C$_{2v}$ unit cells. (a,f) supercell band structure, where the dispersion of the opposite pseudospin states are colored by red and blue. The bulk bands are colored in gray. (b,g) Supercell. (c,h) mode localization calculated at points A and B. (d,e,i,j) Poynting vectors of the pseupospin states.}\label{Fig:Supercells}
\end{figure*}
The multi-objective optimization problem including circular frequency constraints can be written as
\begin{align}
\label{eq:TO}
\nonumber \min_{x_e}\; &\sum_p\alpha_pf_p(x_e)+\sum_q \beta_q h_q(x_e)\\
s.t.\; &\mathbf{K(k)}\mathbf{U}=\omega^2\mathbf{M}\mathbf{U}\\
\nonumber &x_e\in\left[0,1\right].
\end{align}
where $p,q$ are indices representing objective functions involving eigenvectors and eigenvalues, respectively, $\alpha_p$ and $\beta_q$ are weights, $\mathbf{K}$ is the stiffness matrix and $\mathbf{M}$ is the mass matrix. $f_p(x_e)$, which is a function of relative density $x_e$, is only invoked when mode inversion is considered at the third stage of optimization. Eq. (\ref{eq:TO}) is solved deterministically by the gradient based optimizer MMA~\cite{svanberg1987method}. The design sensitivities of the objective functions are calculated using adjoint analysis, with special treatment for repeated eigenvalues and their corresponding eigenvectors~\cite{wu2007note}. The material properties are interpolated using SIMP~\cite{bendsoe1989optimal}. The minimum features of the design are controlled by a filter based on Helmholtz-type partial differential equation where the final design which is free of partial density elements is obtained through Heaviside projection~\cite{lazarov2011filters}.  We show in the SI~\cite{SI} that the optimization problem formulated in Eq. (\ref{eq:TO}) can also generate double Dirac cones under C$_2$ and C$_4$ symmetries, while mimicking the behavior of the corresponding BHZ effective Hamiltonian.  For the C$_{2v}$ and C$_{4v}$ symmetries we focus on here, the minimal design region is specified in Figs. \ref{Fig:schematic}(f,g), with the full unit cells reconstructed through the illustrated symmetry operations. The unit cells are made of aluminum, with shear modulus $\mu=25.556$ GPa and density $\rho=2700$ kg/m$^3$.

We first discuss results for C$_{4v}$ symmetric unit cells in Fig. \ref{Fig:C4v_C2v}(a), where the middle panel shows the unit cell exhibiting double Dirac cones with linear dispersion in the neighborhood of the degeneracy at a normalized frequency $\Omega=1.295$. Here, the two $p$-like modes are degenerate and deterministic, while the two $d$-like modes are accidentally degenerate.  The left panel in Fig. \ref{Fig:C4v_C2v}(a) shows the unit cell for which a complete band gap is opened, lifting the accidental degeneracy between the $p$ and $d$-like modes, while the right panel in Fig. \ref{Fig:C4v_C2v}(a) shows the unit cell after optimization inducing inversion of the modes in the left panel.  The mode inversion process brings the $p$-like modes to the lower edge of the band gap, while the $d$-like modes are brought to the upper edge of the band gap. The complete band gaps within the these two complementary unit cells are centered at the frequency of the original double Dirac cones with upper edge at $\Omega=1.239$ and lower edge at $\Omega=1.351$, resulting in a 8.65\% relative band gap size. 

The results of the optimization process for C$_{2v}$ unit cell are shown in Fig. \ref{Fig:C4v_C2v}(b).  The topologies for the C$_{2v}$ unit cell are quite similar to the C$_{4v}$ unit cells in Fig. \ref{Fig:C4v_C2v}(a), where slight variations of the geometry along the orthogonal directions can be found that demonstrate that the unit cells belong to C$_{2v}$ symmetry group. In this case, both the $p$ and $d$-like modes are accidentally degenerate at either the double Dirac cone or the edges of the topological band gap, where the user defined degenerate frequencies are set to be the same as the ones in C$_{4v}$ cases.  One difference compared to the C$_{4v}$ results is that in opening the band gap in the left panel of Fig. \ref{Fig:C4v_C2v}(b), the optimization brings the degenerate $p$-like modes to the lower edge of the band gap while the $d$-like modes degenerate at the upper edge of the band gap.  Mode inversion inverts the position of the $p$ and $d$-like modes as shown in the right panel of Fig. \ref{Fig:C4v_C2v}(b).  For both the C$_{4v}$ and C$_{2v}$ unit cells, we have verified through calculation of Chern numbers (details in the SI~\cite{SI}) that the unit cells which have quadrupolar $d$-like modes at the lower edge of the topological band gap are topologically non-trivial~\cite{zhang2017experimental}, while the mode inverted counterpart is topologically trivial.  

We verified the existence of opposite pseudospin states by examining the eigenspectral properties of a supercell, which consists of a topologically non-trivial unit cell interfacing with a topologically trivial unit cell, where for consistency the non-trivial unit cells were always below the interface.  As shown by the left panel of Fig. \ref{Fig:Supercells}, the C$_{4v}$ supercell shows two opposite spin states crossing the band gap centered at the degenerate frequency of the double Dirac cones, with a complete bulk band gap matching the design frequencies. There exists a small edge state gap of 0.48\% of the topological band gap, which is caused by the symmetry breaking at the interface~\cite{chenPRB2018,yvesNJP2017}. It is shown in Fig. \ref{Fig:Supercells}(c) that the displacement is localized within one unit cell above and below the interface. The Poynting vector calculations in Figs. \ref{Fig:Supercells}(d,e) show the energy propagation in opposite directions for the two pseudospin states for the C$_{4v}$ case, while similar pseudospin behavior is observed in Figs. \ref{Fig:Supercells}(i,j) for the C$_{2v}$ case.  Because the double Dirac degeneracy for the C$_{2v}$ case originates from two accidental degeneracies, there is a slightly larger edge state gap of 0.98\% of the total band gap.

\begin{figure*}[t]
\center
\includegraphics[width=\textwidth]{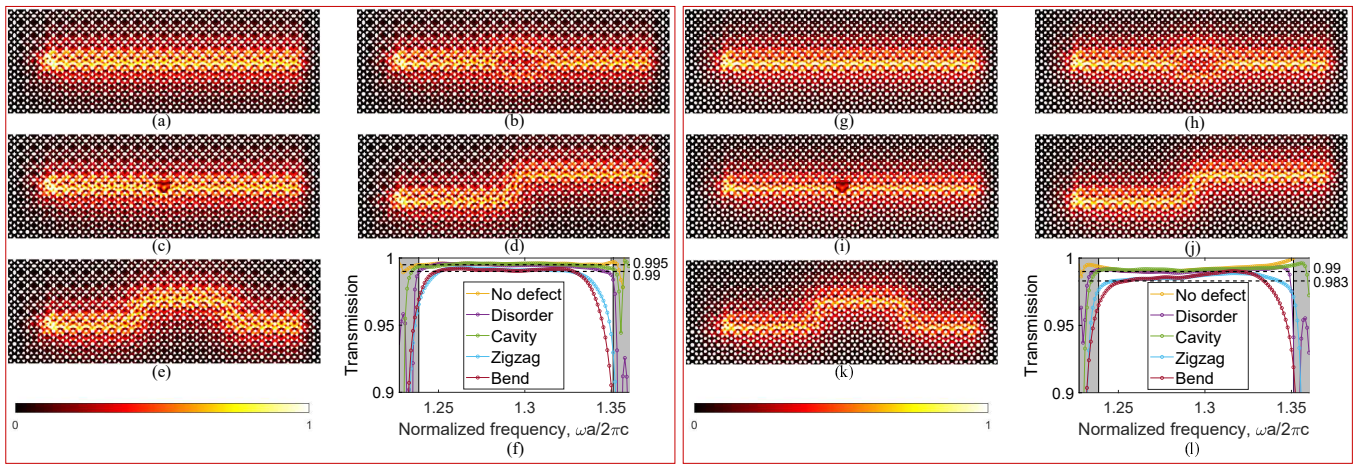}
\caption{One-way wave propagation simulation and transmission calculation for TI constructed from (left panel) C$_{4v}$ and (right panel) C$_{2v}$ unit cells.  (a,g) Defect-free. (b,h) Disorder. (c,i) Defect. (d,j) Zigzag bend. (e,k) Double bend. (f,l) Normalized energy transmission. }\label{Fig:Transmission}
\end{figure*}

Finally, we demonstrate the topologically-protected interfacial wave propagation by presenting a series of one-way wave propagation simulations within the TIs constructed by the designs shown previously. The simulation domain is surrounded by two layers of unit cells working as crystalline perfectly matched layers (PMLs) to minimize scattering at the boundaries. The normalized transmission is measured by comparing the averaged energy flux at the input and output regions of the simulation domain, and is measured within three unit cells above and below the interface, respectively. The simulations are conducted at $\Omega=1.3174$, which is slightly above the pseudospin crossing frequency. The results show that the TIs constructed by optimized C$_{4v}$ and C$_{2v}$ unit cells are immune to defects such as disordered unit cells, cavity, zigzag bend and double bend.  For the C$_{4v}$ cases, the transmission is greater than 0.99 for all cases for the majority of the topological band gap.  For the C$_{2v}$ cases, the transmission is greater than 0.983 for all cases, again showing robustness to disorder and defects.  For the zigzag and double bend cases, the transmission dips slightly at the edges of the band gap in Figs. \ref{Fig:C4v_C2v}(a,b), though for the largest drop in transmission to 0.9, this translates into only a $10\log_{10}0.9=-0.458dB$ loss.  

In conclusion, we report the systematic design of double Dirac cones and topologically non-trivial phonons based on the QSHE for continuous, square symmetric unit cells by a novel gradient-based multi-objective topology optimization approach. The objective functions are judiciously formulated to mimic the eigenmode behaviors of the BHZ effective Hamiltonian, such that the resulting optimized designs exhibit mode degeneracies at user defined frequencies along with matched group velocities, and topologically complementary unit cells that exhibit mode inversion.  Supercell band structure calculations show that the one-way propagating edge modes span the topological band gap for both C$_{4v}$ and C$_{2v}$ cases, leading to robust, defect-immune and back-scattering resistant TIs, where the pseudospin states are formed entirely or in-part by accidental degeneracies through the topology optimization approach. These results thus open the possibility for realizing QSHE-based pseudospin-dependent transport beyond hexagonal lattices.  Finally, because of the universality of the BHZ model for describing the QSHE in photonics and acoustics, our effective Hamiltonian inspired optimization approach can be easily extended to these areas.

HSP and YL acknowledge the support of the Army Research Office, grant W911NF-18-1-0380, and the College of Engineering at Boston University.

\bibliographystyle{apsrev4-1}
%

\end{document}


\title{Supplemental Information for "Double Dirac Cones and Topologically Non-Trivial Phonons for Continuous, Square Symmetric (\texorpdfstring{C$_{4v}$}{Lg} and \texorpdfstring{C$_{2v}$}{Lg}) Unit Cells"}

\author{Yan~Lu}
\affiliation{Department of Mechanical Engineering, Boston University, Boston, MA 02215}

\author{Harold~S.~Park}
\email[Corresponding author: ]{parkhs@bu.edu}
\affiliation{Department of Mechanical Engineering, Boston University, Boston, MA 
02215}

\date{\today}

\maketitle

\section{BHZ Model Details}

The motivation for our design strategy is the QSHE topological system that is described by the Bernevig-Hughes-Zhang (BHZ) model Hamiltonian~\cite{bernevigSCIENCE2006}
\begin{equation}\label{Eq:BHZmodel}
\mathcal{H}(\mathbf{k})=
\begin{bmatrix}
\mathcal{H}_+&0\\
0&\mathcal{H}_-
\end{bmatrix}
\end{equation} 
where
\begin{equation}\label{Eq:BHZmodeldetails}
\mathcal{H}_{\pm}=
\begin{bmatrix}
M+B\delta k^2 & A\delta k_{\mp}\\
A^*\delta k_{\pm} & -M-B\delta k^2
\end{bmatrix}.
\end{equation}
The topological transition between trivial and non-trivial states depends on the behavior of parameters $A$, $B$ and $M$, which determine the strength of the first order mode coupling, second order mode coupling, and the gap size of the topological band gap, respectively. For example, when the two pseudospin states are degenerate, $M$ and $B$ vanish leading to the formation of double Dirac cones. When strong spin-orbit coupling is introduced to lift the four-fold degeneracy, a complete topological band gap is formed with inverted eigenmodes that appear above and below the band gap, while the eigenvalue and group velocity degeneracy is kept intact at the edges of the band gap. The above effective Hamiltonian can be derived from a second order $k\cdot p$ perturbation\cite{dresselhaus2007group,wu2015scheme}. Here, we derive the effective Hamiltonians for C$_{4v}$/C$_{4}$/C$_{2v}$/C$_{2}$ symmetry to show the necessity of our optimization steps as explained in the main text to match the BHZ model. 

The $k\cdot p$ perturbation method is established based on the fact that eigenvectors in the neighborhood of a degeneracy can be expressed as a linear combination of the eigenvectors of the degeneracy, i.e. $\mathbf{u}(\mathbf{k'})=e^{\mathrm{i}\delta \mathbf{k}\cdot \mathbf{r}}\sum_{n=1}^N\psi_n\mathbf{u}_n(\mathbf{k})$, where $N$ is the eigenvalue multiplicity. For a general three-dimensional phononic unit cell, we substitute this relation into the equilibrium equation, $\nabla\cdot[\mathbf{C}:\nabla^s\mathbf{u(k)}]=-E(\mathbf{k})\rho\mathbf{u(k)}$ and get 
\begin{equation}\label{eq:kdotp}
\sum_{n=1}^N\psi_n\left[ E(\mathbf{k})\delta_{mn}+\delta\mathbf{k}\cdot\mathbf{p}+\delta\mathbf{k}\delta\mathbf{k}:\mathbf{q}\right]=E_m(\mathbf{k'})\psi_m,
\end{equation}
where 
\begin{equation}
\label{eq:p}
\mathbf{p}=\langle\mathbf{u}_m|\boldsymbol{p}|\mathbf{u}_n\rangle=-\int \mathrm{i}\mathbf{u}_m^*\cdot\left[\mathbf{C}:\nabla^s+\nabla\cdot(\mathbf{C}\cdot)\right]\mathbf{u}_nd\Omega,
\end{equation}
\begin{equation}
\label{eq:q}
\mathbf{q}=\int \mathbf{u}_m^*\cdot\mathbf{C}\cdot\mathbf{u}_nd\Omega.
\end{equation}
Here, $\mathbf{C}$ is the stiffness tensor and $\mathbf{u}$ is orthonormalized such that $\int\boldsymbol{\rho}\mathbf{u}_m^*\mathbf{u}_n=\delta_{mn}$. Therefore, the effective Hamiltonian can be written as
\begin{equation}
\mathcal{H}=\mathcal{H}_0+\mathcal{H}' ,
\end{equation}
where $\mathcal{H}_0=E(\mathbf{k})\delta_{mn}+\delta\mathbf{k}\delta\mathbf{k}:\mathbf{q}$ and the linear perturbation term $\mathcal{H}'=\delta \mathbf{k}\cdot \mathbf{p}$. $\delta \mathbf{k}$ is the incremental wave number away from the degenerate point.  Subsequently, we will drop the $\delta$ symbol for simplicity. For the anti-plane shear wave motion we consider here, the $m$th displacement mode only contains the out-of-plane component, denoted as $u_m$, and the stiffness tensor is replaced by the shear modulus $\mu$. 

\subsection{\texorpdfstring{C$_{4v}$}{Lg} Unit Cells}

For the double Dirac cones we consider here, $\mathcal{H}_0$ can be written as a diagonal matrix, and $\mathcal{H}'$ can be simplified by checking the non-zero $\langle u_m|\boldsymbol{p}|u_n\rangle$ entries via symmetry analysis. For C$_{4v}$ symmetry the four modes involved, $\mathbf{u}_m$, have the same symmetry as $[p_x,p_y,d_{x^2-y^2},d_{xy}]$, respectively. The $p$-like modes transform as $E$ irreducible representation, $d_{x^2-y^2}$-like mode transform as $B_1$ irreducible representation, $d_{xy}$-like mode transform as $B_2$ irreducible representation, and that operator $\boldsymbol{p}$ transforms as a vector corresponding to $E$ irreducible representation. After checking the direct product of related irreducible representations, we find that the non-zero entries only appear in the off-diagonal blocks and the detailed entries are
\begin{align}\label{eq:firstorder}
\mathcal{H}'_{mn}=\begin{bmatrix}
0& 0& \langle u_1|k_x\boldsymbol{p}_x|u_3\rangle& \langle u_1|k_y\boldsymbol{p}_y|u_4\rangle\\
0& 0& \langle u_2|k_y\boldsymbol{p}_y|u_3\rangle& \langle u_2|k_x\boldsymbol{p}_x|u_4\rangle\\
\langle u_3|k_x\boldsymbol{p}_x|u_1\rangle& \langle u_3|k_y\boldsymbol{p}_y|u_2\rangle& 0& 0\\
\langle u_4|k_y\boldsymbol{p}_y|u_1\rangle& \langle u_4|k_x\boldsymbol{p}_x|u_2\rangle& 0& 0\\
\end{bmatrix}
\end{align}
Notice that there is vanishing $\boldsymbol{p}_x(\boldsymbol{p}_y)$ component of the operator $\boldsymbol{p}$ in C$_{4v}$ symmetry. Evaluation of Eq. (\ref{eq:firstorder}) shows that $\mathcal{H}'_{mn}$ in general has the following form
\begin{align}\label{eq:firstordersimplified}
\mathcal{H}'_{mn}=\begin{bmatrix}
0& 0& Ak_x& Bk_y\\
0& 0& -Ak_y& Bk_x\\
A^*k_x& -A^*k_y& 0& 0\\
B^*k_y& B^*k_x& 0& 0
\end{bmatrix},
\end{align}
where $A$ and $B$ are purely imaginary according to Eq. (\ref{eq:p}). Rewriting Eq. (\ref{eq:firstordersimplified}) on the new basis $[u_1+\mathrm{i}u_2,u_3+\mathrm{i}u_4, u_1-\mathrm{i}u_2,u_3-\mathrm{i}u_4]$ with $k_{\pm}=k_x\pm\mathrm{i}k_y$, we have
\begin{equation}\label{eq:firstorderspin}
\mathcal{H}'_{mn}=\begin{bmatrix}
0& \dfrac{(A+B)k_-}{2}& 0& \dfrac{(A-B)k_-}{2}\\
\dfrac{(A^* + B^*)k_+}{2}& 0& \dfrac{(A^* - B^*)k_-}{2}& 0\\
0& \dfrac{(A - B)k_+}{2}& 0& \dfrac{(A + B)k_+}{2}\\
\dfrac{(A^* - B^*)k_+}{2}& 0& \dfrac{(A^* + B^*)k_-}{2}& 0 
\end{bmatrix}
\end{equation}
Eq. (\ref{eq:firstorderspin}) gives two sets of eigenvalues $\pm A|\mathbf{k}|$ and $\pm B|\mathbf{k}|$, which describe double Dirac cones with unmatched eigenvalue surfaces. However, the BHZ model demands that the eigenvalue surfaces should also match, essentially, $A=B$. It, therefore, necessitates the topology optimization step which seeks to minimize the eigensurface difference such that the effective Hamiltonian will take the following block diagonal form
\begin{equation}\label{eq:firstordersurfacematched}
\mathcal{H}'_{mn}=\begin{bmatrix}
0& Ak_- & 0& 0\\
A^*k_+& 0& 0& 0\\
0& 0& 0& Ak_+\\
0& 0& A^*k_-& 0 
\end{bmatrix},
\end{equation}

When a topological band gap is opened, the above effective Hamiltonian will no longer be adequate in describing the system, because now it must involve second order perturbation to lift the degeneracy, where the formulation is given by \cite{dresselhaus2007group}
\begin{equation}\label{eq:secondorderformulation}
\mathcal{H}'_{mn}\rightarrow\mathcal{H}'_{mn}+\sum_\alpha\dfrac{\mathcal{H}'_{m\alpha}\mathcal{H}'_{\alpha n}}{E_m-E_{\alpha}}.
\end{equation}
where $\alpha$ labels the degenerate mode on the other edge of the gap opposite to the $m$th mode. The two sets of degenerate eigenvalues are given as $[E_p,E_p,E_d,E_d]$ for their $p$- and $d$-like modes. Due to the unbroken C$_{4v}$ symmetry during the optimizations, components of $\mathcal{H}'_{m\alpha}$ and $\mathcal{H}'_{\alpha n}$ of the optimized unit cells have the same form as in Eqs. (\ref{eq:firstorder}) and (\ref{eq:firstordersimplified}). The effective Hamiltonian for the general C$_{4v}$ case can now be written as 
\begin{equation}\label{eq:secondorder}
\begin{bmatrix}
E_p+\dfrac{|A|^2k_x^2+|B|^2k_y^2}{E_p-E_d}& -\dfrac{|A|^2-|B|^2}{E_p-E_d}k_xk_y & Ak_x& Bk_y\\
 -\dfrac{|A|^2-|B|^2}{E_p-E_d}k_xk_y& E_p+\dfrac{|A|^2k_y^2+|B|^2k_x^2}{E_p-E_d}& -Ak_y& Bk_x\\
 A^*k_x& -A^*k_y& E_d-\dfrac{|A|^2(k_x^2+k_y^2)}{E_p - E_d}& 0\\
 B^*k_y& B^*k_x& 0& E_d-\dfrac{|B|^2(k_x^2+k_y^2)}{E_p - E_d}
\end{bmatrix}
\end{equation}

Rewriting Eq. (\ref{eq:secondorder}) using the pseudospin states as the new basis $[u_1+\mathrm{i}u_2,u_3+\mathrm{i}u_4, u_1-\mathrm{i}u_2,u_3-\mathrm{i}u_4]$, we have
\begin{equation}\label{eq:secondorderspin}
\begin{bmatrix}
E_p+\dfrac{(|A|^2+|B|^2)k^2}{2(E_p-E_d)}& \dfrac{(A+B)k_-}{2}& \dfrac{(|A|^2 - |B|^2)k_-^2}{2(E_p - E_d)}& \dfrac{(A - B)k_-}{2}\\
\dfrac{(A^* + B^*)k_+}{2}& E_d-\dfrac{(|A|^2+|B|^2)k^2}{2(E_p-E_d)}& \dfrac{(A^* - B^*)k_-}{2}& -\dfrac{(|A|^2 - |B|^2)k^2}{2(E_p - E_d)}\\
\dfrac{(|A|^2 - |B|^2)k_+^2}{2(E_p - E_d)}& \dfrac{(A - B)k_+}{2}& E_p+\dfrac{(|A|^2+|B|^2)k^2}{2(E_p-E_d)}& \dfrac{(A + B)k_+}{2}\\
\dfrac{(A^* - B^*)k_+}{2}& -\dfrac{|A|^2 - |B|^2)k^2}{2(E_p - E_d)}& \dfrac{(A^* + B^*)k_-}{2}& E_d-\dfrac{(|A|^2+|B|^2)k^2}{2(E_p-E_d)}
\end{bmatrix},
\end{equation}
where $k_{\pm}=k_x\pm\mathrm{i}k_y$. Observe that the matrix entries of Eq. (\ref{eq:firstorderspin}) now appear at the same location in Eq. (\ref{eq:secondorderspin}). Similar to Eq. (\ref{eq:firstorderspin}), this effective Hamiltonian gives two sets of unmatched eigenvalues surfaces which are degenerate only at the high symmetry ($M$) point mentioned in the main text. Therefore, we again leverage topology optimization to minimize the frequency difference between eigenvalue surfaces such that effectively $A=B$ and the Hamiltonian can be reduced to
\begin{equation}\label{eq:secondordersurfacematched}
\begin{bmatrix}
E_p+\dfrac{|A|^2k^2}{E_p-E_d}& Ak_-& 0& 0\\
A^*k_+& E_d-\dfrac{|A|^2k^2}{E_p-E_d}& 0& 0\\
0& 0& E_p+\dfrac{|A|^2k^2}{E_p-E_d}& Ak_+\\
0& 0& A^*k_-& E_d-\dfrac{|A|^2k^2}{E_p-E_d}
\end{bmatrix}
\end{equation}
Finally, we let $M=(E_p-E_d)/2$ and $B=|A|^2/(E_p-E_d)+D$, where $D$ comes from the second order term in $\mathcal{H}_0$. This brings the effective Hamiltonian to the block diagonal form as in Eqs. (\ref{Eq:BHZmodel}) and (\ref{Eq:BHZmodeldetails}).

\subsection{\texorpdfstring{C$_{2v}$}{Lg} Unit Cells}

For C$_{2v}$ unit cells with double Dirac cones, symmetry analysis show that the effective Hamiltonian has the same form with Eq. (\ref{eq:firstorder}), where there is vanishing operator components $\boldsymbol{p}_x(\boldsymbol{p}_y)$. However, due to the lack of C$_4$ symmetry, $\mathcal{H}'_{mn}$ has the following form
\begin{align}\label{eq:firstordersimplifiedC2v}
\mathcal{H}'_{mn}=\begin{bmatrix}
0& 0& Ak_x& Bk_y\\
0& 0& -Ck_y& Dk_x\\
A^*k_x& -C^*k_y& 0& 0\\
B^*k_y& D^*k_x& 0& 0
\end{bmatrix}.
\end{align}
Again, Eq. (\ref{eq:firstordersimplifiedC2v}) corresponds to two sets of unmatched eigenvalues surfaces. Furthermore, the group velocity is anisotropic due to the eigenvalues' dependence upon both $k_x$ and $k_y$. We use topology optimization to effectively minimize the frequency differences between the eigenvalues surfaces, such that the effective Hamiltonian of the resultant unit cell will give $A=B=C=D$, and take the block diagonal form as in Eq. (\ref{eq:firstorderspin}).
For the second order perturbation, we substitute Eq. (\ref{eq:firstordersimplifiedC2v}) into Eq. (\ref{eq:secondorderformulation}) and change basis from $u_m$ to $[u_1+\mathrm{i}u_2,u_3+\mathrm{i}u_4, u_1-\mathrm{i}u_2,u_3-\mathrm{i}u_4]$ to get $\mathcal{H}'$. The block diagonal form will hold when the eigenvalues surfaces are matched with $A=B=C=D$.

\subsection{\texorpdfstring{C$_4$}{Lg} and \texorpdfstring{C$_2$}{Lg} Unit Cells}

For C$_4$ and C$_2$ unit cells with double Dirac cones, there is no vanishing operator component of $\boldsymbol{p}$, therefore,
\begin{equation}\label{eq:firstorderC4_C2}
\mathcal{H}'_{mn}=\begin{bmatrix}
0& 0& \langle u_1|\mathbf{k}\cdot\boldsymbol{p}|u_3\rangle& \langle u_1|\mathbf{k}\cdot\boldsymbol{p}|u_4\rangle\\
0& 0& \langle u_2|\mathbf{k}\cdot\boldsymbol{p}|u_3\rangle& \langle u_2|\mathbf{k}\cdot\boldsymbol{p}|u_4\rangle\\
\langle u_3|\mathbf{k}\cdot\boldsymbol{p}|u_1\rangle& \langle u_3|\mathbf{k}\cdot\boldsymbol{p}|u_2\rangle& 0& 0\\
\langle u_4|\mathbf{k}\cdot\boldsymbol{p}|u_1\rangle& \langle u_4|\mathbf{k}\cdot\boldsymbol{p}|u_2\rangle& 0& 0\\
\end{bmatrix}
\end{equation}
Specifically, for C$_4$ and C$_2$ 
\begin{align}\label{eq:firstordersimplifiedC4}
\mathcal{H}'_{mn}=\begin{bmatrix}
0& 0& A_1k_x-A_2k_y& B_1kx+B_2k_y\\
0& 0& -A_2k_x-A_1k_y& B_2k_x-B_1k_y\\
A_1^*k_x-A_2^*k_y& -A_2^*k_x-A_1^*k_y& 0& 0\\
B_1^*kx+B_2^*k_y& B_2^*k_x-B_1^*k_y& 0& 0
\end{bmatrix}_{(C_4)},
\end{align}
\begin{align}\label{eq:firstordersimplifiedC2}
\mathcal{H}'_{mn}=\begin{bmatrix}
0& 0& A_1k_x+A_2k_y& -B_1kx+B_2k_y\\
0& 0& C_1k_x-C_2k_y& D_1k_x+D_2k_y\\
A_1^*k_x+A_2^*k_y& C_1^*k_x-C_2^*k_y& 0& 0\\
-B_1^*kx+B_2^*k_y& D_1^*k_x+D_2^*k_y& 0& 0
\end{bmatrix}_{(C_2)}.
\end{align}
For C$_4$ symmetry, the two sets of eigenvalues, $\pm V_{\pm}|\mathbf{k}|$, again describe double Dirac cones with unmatched eigenvalue surfaces. Since the eigenvalues only depend on $|\mathbf{k}|$, the group velocities are isotropic and given as
\begin{align}
V_{\pm}=&\dfrac{\sqrt{2(|A_1|^2+|A_2|^2+|B_1|^2+|B_2|^2\pm\Delta)}}{2},\\
\nonumber \Delta=&|A_1|^4 + 2|A_1|^2|A_2|^2 + 2|A_1|^2|B_1|^2 - 2|A_1|^2|B_2|^2- 4A_1A_2^*B_2B_1^* + |A_2|^4 - 4A_2A_1^*B_1B_2^*\\
\nonumber &- 2|A_2|^2|B_1|^2 + 2|A_2|^2|B_2|^2 + |B_1|^4 + 2|B_1|^2|B_2|^2 + |B_2|^4.
\end{align} 
For C$_2$ symmetry, the eigenvalues' dependence on both $k_x$ and $k_y$ leads to anisotropic group velocity, and the eigenvalue surfaces are also unmatched. We then substitute Eq. (\ref{eq:firstordersimplifiedC4}) and Eq. (\ref{eq:firstordersimplifiedC2}) into Eq. (\ref{eq:secondorderformulation}) and change to the new basis to obtain the effective Hamiltonians under second order perturbation for C$_4$ and C$_2$ respectively. The block diagonal form will hold when the eigenvalue surfaces are matched, while the anisotropy in group velocity is addressed by topology optimization.
 
\subsection{Comparison of BHZ Effective Hamiltonians and Finite Element Simulations}

\begin{figure*}
\includegraphics[width=\textwidth]{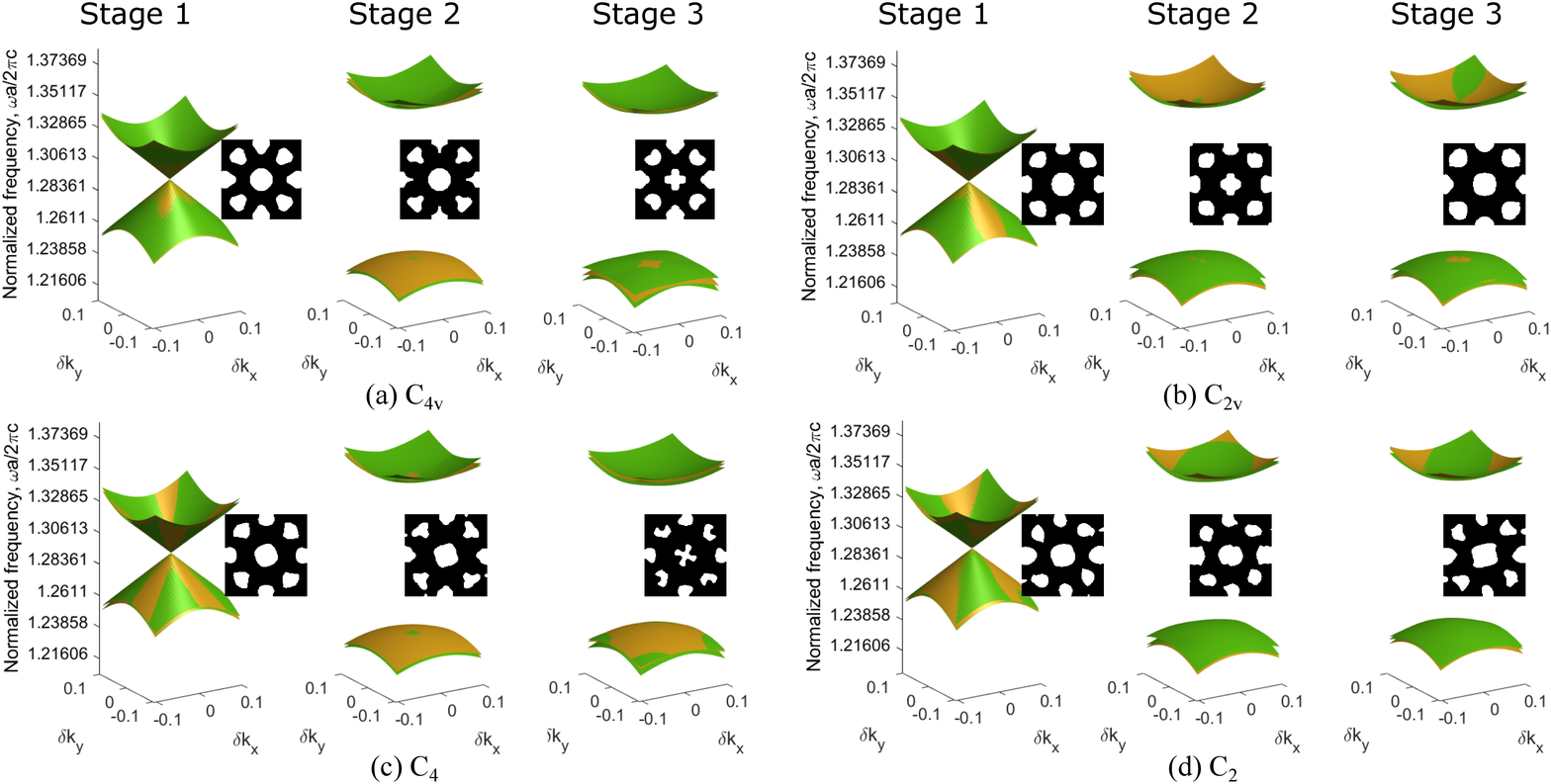}
\caption{Eigenvalue surfaces of optimized unit cells with (a) C$_{4v}$ symmetry, (b) C$_{2v}$ symmetry, (c) C$_{4}$ symmetry, (b) C$_{2}$ symmetry. The surfaces which are calculated using FEM are colored in green, and the surfaces which are calculated using effective Hamiltonian are colored in yellow.}\label{Fig:surfacematching}
\end{figure*}
The finite element calculations of the optimized C$_{4v}$/C$_4$/C$_{2v}$/C$_2$ unit cells and the eigenvalues from the above perturbation theory have shown that the effective Hamiltonian not only matches the numerical results, but also successfully approximates the BHZ model due to the optimization of the frequency difference between the surfaces, as shown in Fig. \ref{Fig:surfacematching}. From the previous section, we can see that the difficulty in designing C$_{4v}$/C$_4$ symmetric cases lies in matching the eigenvalues surfaces, while for C$_{2v}$/C$_2$ symmetric cases, it lies in both the cancellation of anisotropy of group velocity and the matching of eigenvalue surfaces. Due to the fact that the minimal design domains for C$_4$ and C$_{2v}$ symmetric cases are of the same size which occupies a quarter of the square lattice, these two cases have roughly the same design complexity for the optimizer when it searches for the optimal configuration in the feasible region. However, the minimal design domain for the C$_2$ symmetric case occupies one half of the square lattice, which significantly increases the design complexity, therefore, it is in general more difficult to generate structures that satisfy the conditions set forth by the BHZ effective Hamiltonian. 

\subsection{Chern Number Calculations}

\begin{figure}
\includegraphics[scale=0.3]{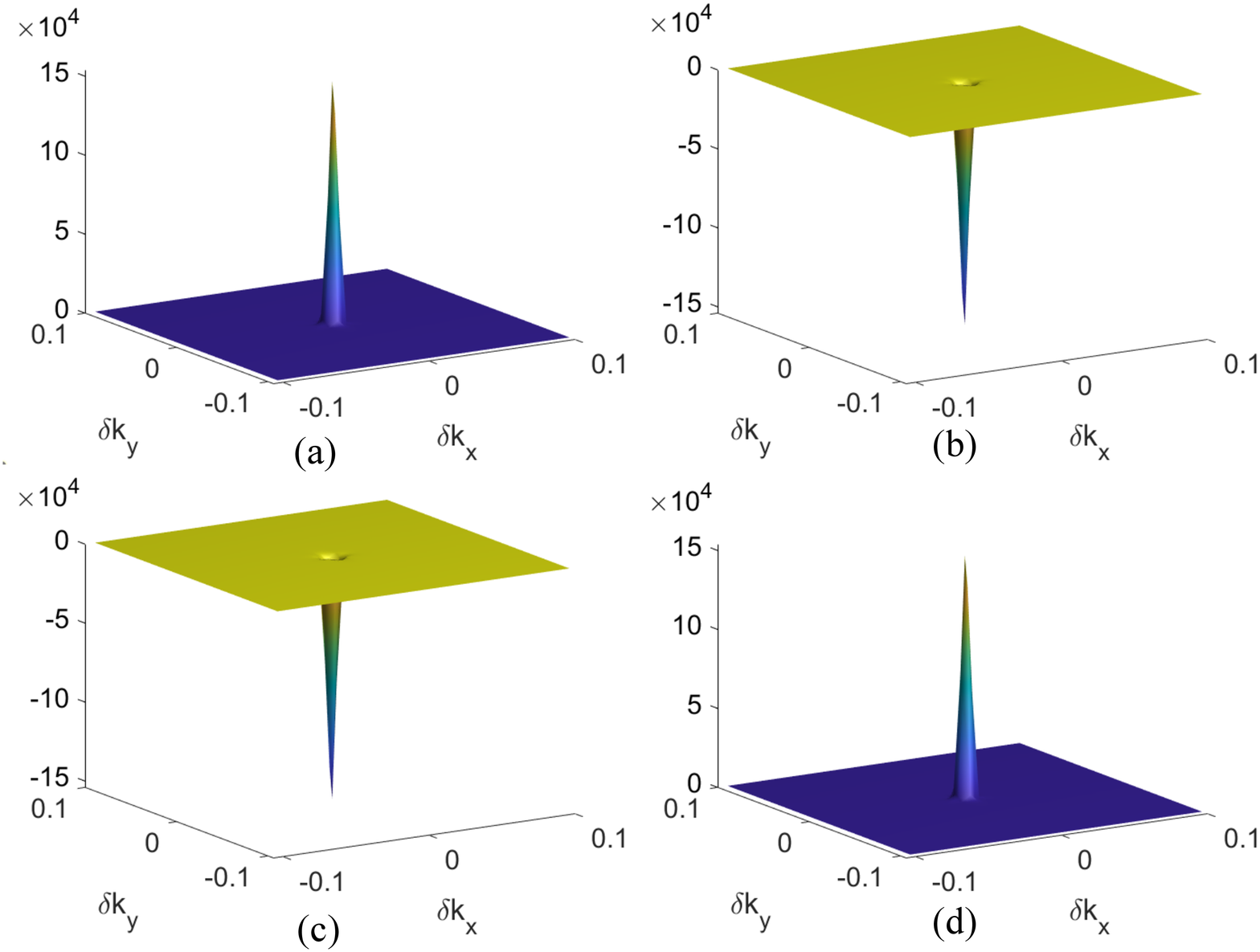}
\caption{Berry curvature for C$_{4v}$ unit cell corresponding to Fig. 2(a) left panel in the main text. (a)(b) Berry curvature belongs to the two lower bands around the $M$ point.(c)(d) Berry curvature belongs to the two upper bands around the $M$ point.}\label{Fig:Berrycurvature}
\end{figure}

To verify that we indeed generate topologically non-trivial unit cells, we numerically calculated the Berry curvature and spin Chern number via the $k\cdot p$ perturbation approach. The details of the Chern number calculations have been given in \cite{wu2015scheme}. As an example, we show in Fig. \ref{Fig:Berrycurvature} the Berry curvature of the C$_{4v}$ unit cell corresponding to the left panel of Fig. 2(a) in the main text. The non-zero zero Berry curvature occurs near the $M$ point. Integrating the Berry curvature over the first Brillouin zone gives $C_s=\pm1$ for the non-trivial unit cell. The Chern number calculations for C$_4$/C$_{2v}$/C$_2$ unit cells with topological band gaps have shown that the unit cells which give $d$-like modes below the $p$-like modes are non-trivial, while the ones which give $p$-like modes below the $d$-like modes are trivial. 

\section{Topology optimization}

\begin{figure*}[b]
\includegraphics[width=\textwidth]{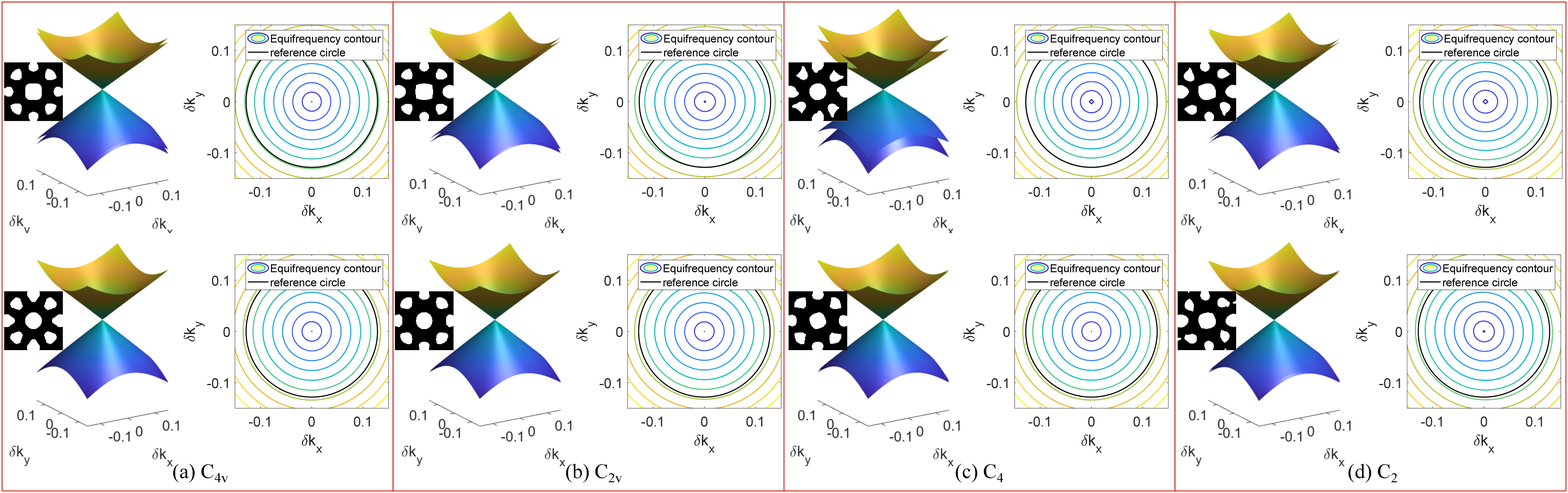}
\caption{Eigenvalues surfaces and equifrequency contours of optimized unit cells with (a) C$_{4v}$ symmetry, (b) C$_{2v}$ symmetry, (c) C$_{4}$ symmetry, (b) C$_{2}$ symmetry. The upper panel in each subfigure results from the optimization where the circular frequency constraints are not invoked, and the lower panel results from the optimization where the circular frequency constraints are invoked.  The eigenvalue surfaces and contours are calculated using the $k\cdot p$ method.}\label{Fig:Circularconstraints}
\end{figure*}
As discussed in the main text, the topology optimization consists of three stages. In this SI, we give more details on the set of objective functions which seek to approximate the coalescence of eigenvalue surfaces and isotropy of group velocity.
\begin{align}
\label{eq:isotropicVelocity}
&h=|\max_{l}\lbrace\omega(\mbf{k+\Delta k}_l)_i\rbrace - \min_{l}\lbrace\omega(\mbf{k+\Delta k}_l)_i\rbrace|,\\ 
\label{eq:slopeMatchLowerBraches}
&h=|\max_{l}\lbrace\omega(\mbf{k+\Delta k}_l)_{\ut{i}_2}\rbrace - \min_{l}\lbrace\omega(\mbf{k+\Delta k}_l)_{\ut{i}_1}\rbrace|,\\
\label{eq:slopeMatchupperBraches}
&h=|\max_{l}\lbrace\omega(\mbf{k+\Delta k}_l)_{\tilde{i}_2}\rbrace - \min_{l}\lbrace\omega(\mbf{k+\Delta k}_l)_{\tilde{i}_1}\rbrace|,\\
\nonumber &l=1,\cdots 8.
\end{align}
\Eqref{eq:isotropicVelocity} concerns the isotropy of the dispersion relation of the 4 modes involved. For the $i$th band, eigenvalues are calculated along a small circle of radius $|\mbf{\Delta k}_l|=0.1|\Gamma X|$ centered at $\mbf{k}$. We let $l$ denote the 8 points along the smaller circle shown in Fig. 1(e) in the main text. The absolute values of the differences between the maximal and minimal frequencies on this circle are then optimized, such that the equifrequency contours of the cones become circular. With Eqs. (\ref{eq:isotropicVelocity}) and (1) in the main text, we will have two circular double cones centered at $\mbf{k}$ with coalescing Dirac frequency $\omega_0$. However, the slopes of the two double cones are not guaranteed to match as shown in Eq. (\ref{eq:firstorderspin}). To resolve this, we introduce \Eqref{eq:slopeMatchLowerBraches} and \Eqref{eq:slopeMatchupperBraches}. \Eqref{eq:slopeMatchLowerBraches} considers the two branches, marked by subscripts $\ut{i}_1$ and $\ut{i}_2$, below the Dirac frequency. Here, we assume that the band index $\ut{i}_2$ is larger than $\ut{i}_1$. We find the maximum frequency along the circle of branch $\ut{i}_2$ and minimum frequency along the circle of branch $\ut{i}_1$ and optimize the difference. When the vertices of the cone are fixed at $\omega_0$ and the dispersion relation is approximately linear in the vicinity of $\mbf{k}$, \Eqref{eq:slopeMatchLowerBraches} essentially minimizes the differences between the two conical surfaces. \Eqref{eq:slopeMatchupperBraches} acts upon the two branches above the Dirac frequency that is marked by subscripts $\tilde{i}_1$ and $\tilde{i}_2$, where $\tilde{i}_2>\tilde{i}_1$, for the purpose of eigenvalue surface matching.  The above set of objective functions are also invoked during the second and third stage of optimization, again to ensure coalescence of the eigenvalue surfaces at the top and bottom edges of the topological bandgap. 

\begin{figure*}
\includegraphics[width=\textwidth]{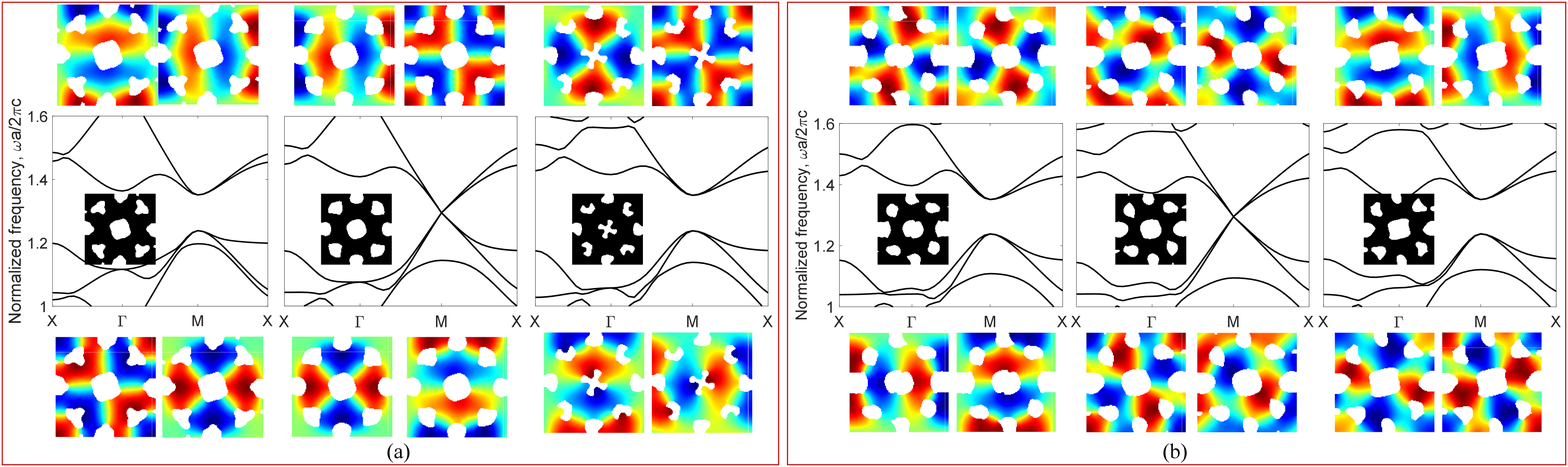}
\caption{(a) Optimized C$_{4}$ unit cells that exhibit double Dirac cones (middle), band gap opening in second stage of optimization (left) and mode inversion (right).(b) Optimized C$_{2}$ unit cells that exhibiting double Dirac cones (middle), band gap opening in second stage of optimization (left) and mode inversion (right).}\label{Fig:C4_C2}
\end{figure*}
To see the effectiveness of the circular frequency constraints, we compare the results with a new set of unit cells of different symmetries with double Dirac cones which are generated by dropping the circular frequency constraints. As shown in the upper panels of Fig. \ref{Fig:Circularconstraints}, which are optimization results when Eqs.(\ref{eq:isotropicVelocity}), (\ref{eq:slopeMatchLowerBraches}) and (\ref{eq:slopeMatchupperBraches}) are not invoked, all the eigenvalue surfaces show splitting. The equifrequency contours for C$_{4v}$/C$_4$ unit cells are circular, and the ones for C$_{2v}$/C$_2$ unit cells are elliptic. In contrast, as shown in the lower panels of Fig. \ref{Fig:Circularconstraints}, the coalescence of the eigenvalues surfaces is realized when the circular frequency constraints are invoked. The roundness of the equifrequency contour is also improved for C$_{2v}$/C$_2$ unit cells. Additional results using C$_4$ and C$_2$ symmetry are shown in Fig. \ref{Fig:C4_C2}, where band structure and modeshapes are calculated. 

\begin{figure*}
\includegraphics[width=\textwidth]{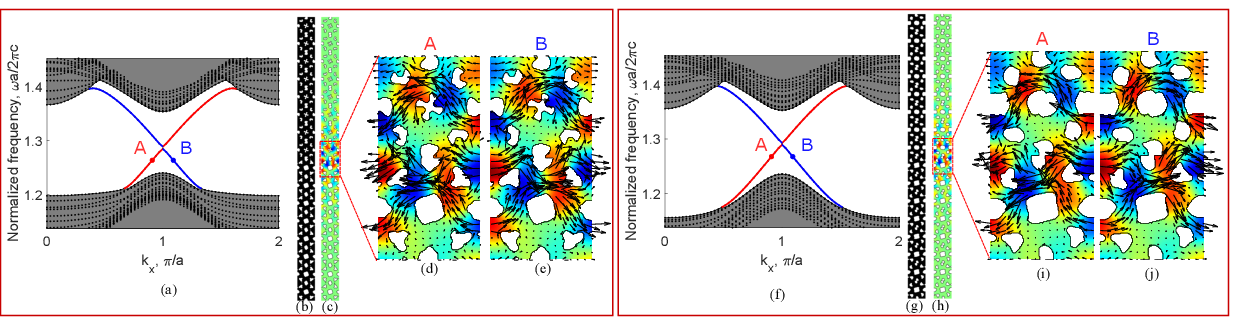}
\caption{Band structure calculations for supercell constructed from (left) C$_{4}$ and (right) C$_{2}$ unit cells. (a,f) supercell band structure, where the dispersion of the opposite pseudospin states are colored by red and blue. The bulk bands are colored in gray. (b,g) Supercell. (c,h) mode localization calculated at points A and B. (d,e,i,j) Poynting vectors of the pseupospin states.}\label{Fig:C4C2Supercells}
\end{figure*}
The super cell band structures and edge states for C$_4$ and C$_2$ are shown in Fig. \ref{Fig:C4C2Supercells}, where opposite spin states cross the band gap, with a complete bulk band gap matching the design frequencies as in Fig. \ref{Fig:C4C2Supercells}(a,f). There exist a small edge state gap of 3.7\% of the topological band gap in C$_4$ super cell, and we also found a similar edge state gap of 2.9\% in C$_2$ super cell. It is shown in both panels of Fig. \ref{Fig:C4C2Supercells} that the displacement localizes within one unit cell above and below the interface, and the energy propagates in opposite directions for the pseudospins in different states. 

\begin{figure*}
\includegraphics[width=\textwidth]{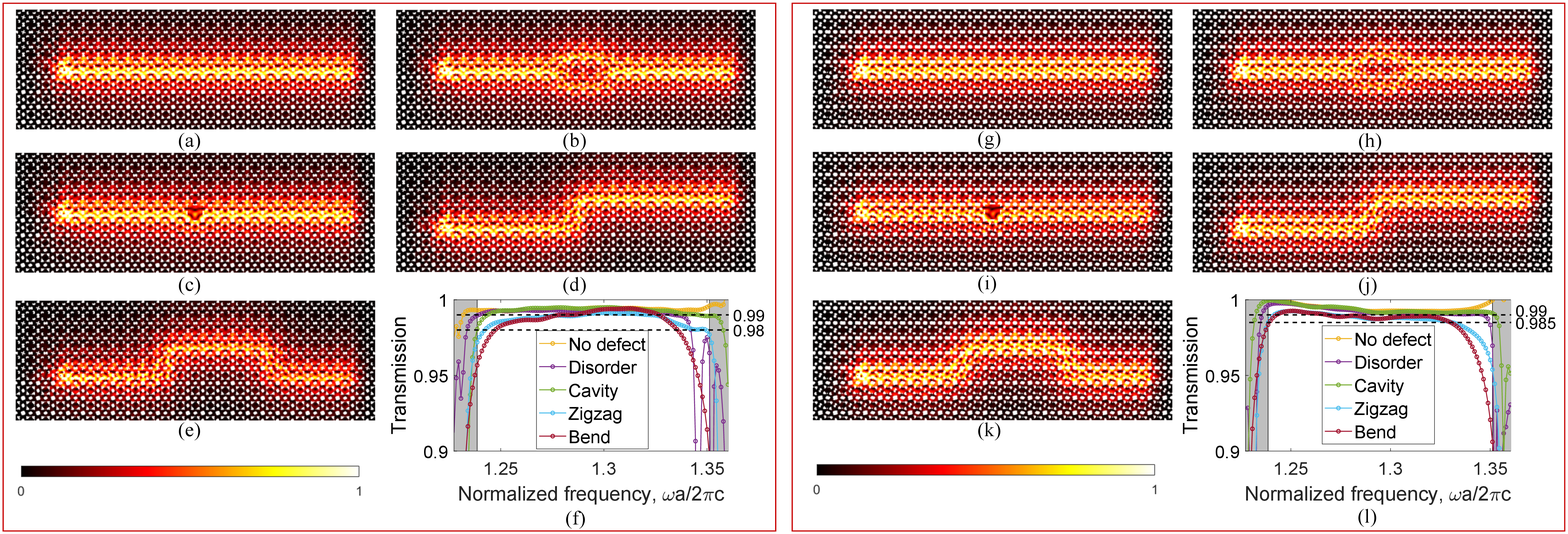}
\caption{One-way wave propagation simulation and transmission calculation for TI constructed from (left panel) C$_4$ and (right panel) C$_{2}$ unit cells.  (a,g) Defect-free. (b,h) Disorder. (c,i) Defect. (d,j) Zigzag bend. (e,k) double bend. (f,l) Normalized energy transmission. }\label{Fig:C4C2Transmission}
\end{figure*}

Finally, the one-way wave propagation simulation and transmission results are shown in Fig. \ref{Fig:C4C2Transmission}. The one-way wave propagation simulation is again done at $\Omega=1.3174$. It is seen that the TIs constructed by C$_4$ and C$_2$ unit cells can resist defects such as disorder, cavity, zigzag bend and double bend. However, due to the existence of the small edge state gap, the transmission of zigzag and double bend cases in C$_4$ falls below 99\% below the frequency where the small gap occurs, and there is a sharp transmission drop for the disorder case near the upper edge of the topological band gap. For the C$_2$ unit cell, the transmission is consistently above 98.5\% for the majority of the gap. 

\begin{figure}
\includegraphics[scale=.55]{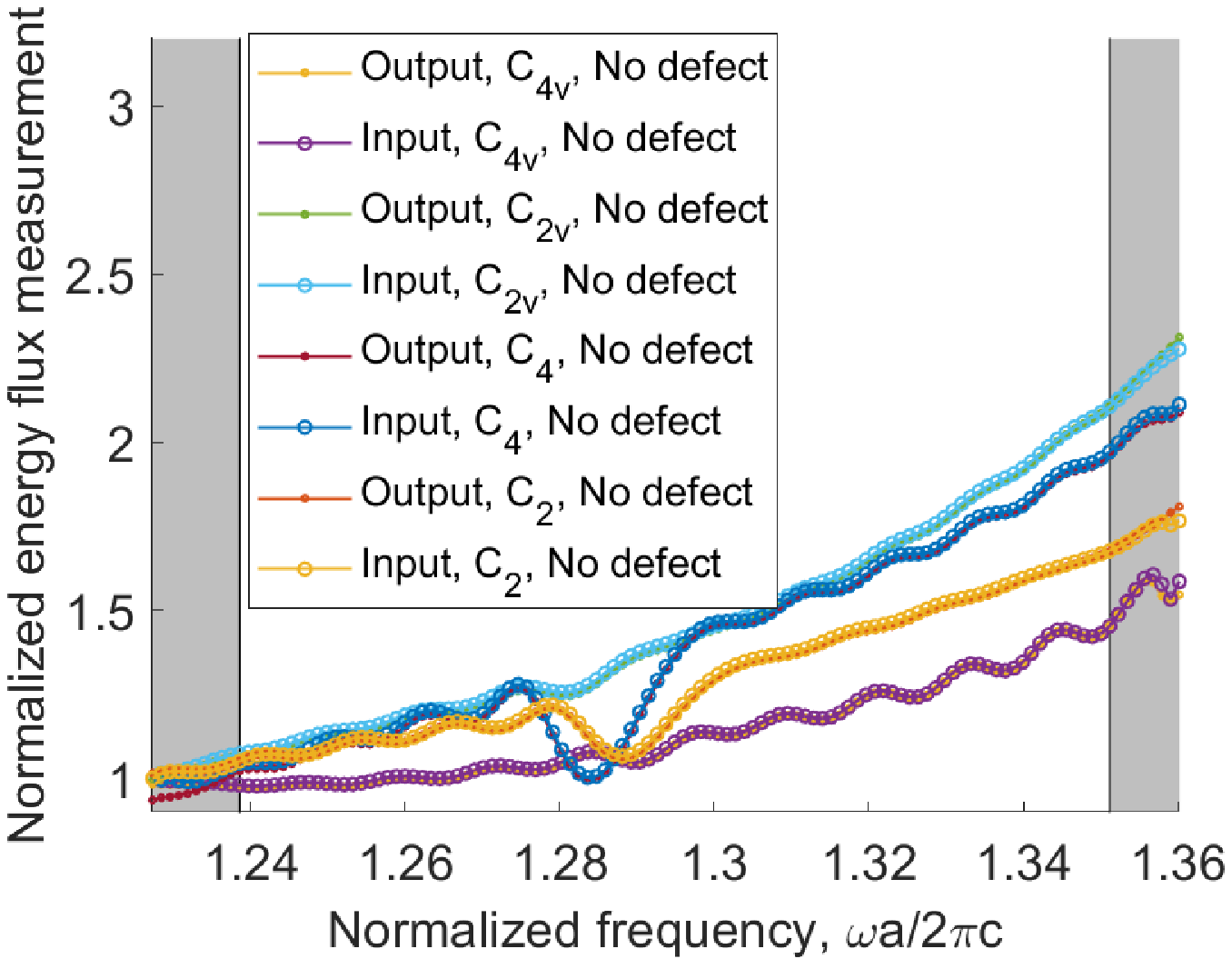}
\caption{The averaged energy flux measured at the input and output measuring area for different frequencies. The results are normalized with respect to the averaged input energy flux at the lowest frequency measured.}\label{Fig:Energymeasure}
\end{figure}

The energy transmission is calculated by taking the ratio between the averaged output and input energy flux. The input is measured within an area which is three unit cells away from the source, similar to \cite{yu2018elastic}, and the output is measured near the right boundary of the simulation domain, which is fourteen unit cells away from the input measuring area. As shown in Fig. \ref{Fig:Energymeasure}, there is a significant drop of input energy flux value for the C$_4$ and C$_2$ cases at the edge state gaps, which implies that due to the edge state gap, not all of the input energy from the source is directed into the topologically-protected interface.  However, the energy that does reach the interface is well-protected, which is why the normalized transmission is still close to 1, as shown in Figs. \ref{Fig:C4C2Transmission}(f,i).  On the other hand, the edge state gaps are much smaller for the C$_{4v}$ and C$_{2v}$ cases, which show no clear energy drop at the edge state crossings in Fig. \ref{Fig:Energymeasure}, which results in superior transmission results as compare to the C$_{4}$ and C$_{2}$ cases.

\bibliographystyle{apsrev4-1}
%